\documentclass{xarticle}

\usepackage[rgb]{xcolor}
\usepackage{amsfonts,amssymb,amsmath,amsthm}
\usepackage{graphicx}
\usepackage{overpic}
\usepackage[shortlabels]{enumitem}

\def\clap#1{\hbox to 0pt{\hss#1\hss}}
\newcommand{\Z}{\mathbb{Z}}
\newcommand{\R}{\mathbb{R}}
\newtheorem{thm}{Theorem}

\newtheorem{lem}[thm]{Lemma}

\def\qed{\rule[0pt]{5pt}{5pt}\par\medskip}
\newcommand{\floor}[1]{\lfloor#1\rfloor}\let\mathopfont=\mathrm

\newcommand{\Null}{\mathop{\mathopfont{null}}}
\def\tp{\mathsf{T}}
\newcommand{\bmat}[1]{\begin{bmatrix}#1\end{bmatrix}}
\def\one{\mathbf{1}}
\let\bl\bigl
\let\br\bigr
\newcommand{\half}{\frac{1}{2}}
\newcommand{\norm}[1]{\lVert{#1}\rVert}
\newbox\vcbox
\def\vcent#1{\setbox\vcbox\hbox{#1}\raise -0.5\ht\vcbox\hbox{#1}}

\usepackage[margin=10pt,font=small,skip=7pt]{caption}
\usepackage[numbers,sort]{natbib}
\def\prerefspace{\vspace*{-0.5ex}}
\def\precapspace{\captionsetup{margin=30pt,font=small,skip=17pt}}

\def\ugn{\lambda}
\newcommand{\neighbors}{\mathop{\mathopfont{neighbors}}}
\def\epoch{t^\text{e}}
\def\wu{\omega^\text{u}}
\def\wc{\omega^\text{c}}
\def\kp{a}
\def\wcki{b}
\def\wavg{\omega^\text{avg}}
\def\twonorm{$2$-norm}
\newcommand{\bittide}{{bittide\ }}
\def\itoj{{ij}}
\def\jtoi{{ji}}

\begin{document}

\title{Resistance Distance and Control Performance for\\
  \bittide Synchronization}

\author{Sanjay Lall\footnotesymbol{1}
  \and  C\u{a}lin Ca\c{s}caval\footnotesymbol{2}
  \and  Martin Izzard\footnotesymbol{2}
  \and  Tammo Spalink\footnotesymbol{2}}

\note{Preprint}

\maketitle

\makefootnote{1}{S. Lall is with the Department of Electrical
  Engineering at Stanford University, Stanford, CA 94305, USA, and is
  a Visiting Researcher at Google.
  \texttt{lall@stanford.edu}\medskip}

\makefootnote{2}{C\u{a}lin Ca\c{s}caval, Martin Izzard, and Tammo Spalink are
  with Google.}

\begin{abstract}
  We discuss control of \bittide distributed systems, which are
  designed to provide logical synchronization between networked
  machines by observing data flow rates between adjacent systems at
  the physical network layer and controlling local reference clock
  frequencies. We analyze the performance of approximate
  proportional-integral control of the synchronization mechanism and
  develop a simple continuous-time model to show the resulting
  dynamics are stable for any positive choice of gains. We then
  construct explicit formulae to show that closed-loop performance
  measured using the $L_2$ norm is a product of two terms, one
  depending only on resistance distances in the graph, and the other
  depending only on controller gains.
\end{abstract}

\section{Introduction}

In this paper, we discuss control of the \bittide synchronization
mechanism for distributed computing.  The purpose of this mechanism is
to provide all of the machines on a network with shared logical
time.  This notion of time does not have to match physical
time. Instead, the discrete clocks of the machines on the network are
tied together in \emph{logical synchronization}.  This is distinct from
\emph{physical synchronization}, where the processor clocks are kept
synchronized to physical
time~\cite{ntp,corbett_spanner_2013}. Application processes on this
system coordinate their actions using logical time, and do not need to
reference physical time.  The \bittide control system
maintains perfect logical synchronization using imperfect physical
synchronization.

We view the network as an undirected graph where each edge represents
a pair of data links between nodes (machines), one link in each
direction.  The synchronization mechanism operates at the physical
layer of the network as follows.  Frames of data received from each
link are appended at the tail of per-link queues called \emph{elastic
buffer}s.  One frame is removed from the head of each elastic buffer
every local clock cycle and consumed by the corresponding compute
core.  One frame is also transmitted on every outgoing link during
every cycle.  This alignment of receive and transmit at each node
reveals the relative frequency between neighboring nodes. If the
elastic buffer at a node starts to fill up, then it must have a lower
clock frequency than its neighbor and vice versa.

For a network with $n$ nodes, if node~$i$ has~$d_i$ neighbors, then it
has $d_i$ elastic buffers, one per node. Frames are removed from all
of the~$d_i$ elastic buffers simultaneously, driven by the local
clock.  At each node~$i$, the local clock is driven by a physical
oscillator, with uncorrected frequency~$\wu_i$.  The uncorrected
frequencies at nodes will differ slightly in practice, and so
additional correction is necessary to ensure system stability. Each
node includes a feedback control system which measures the occupancy
of all local elastic buffers and adds a correction $c_i$ to the local
oscillator frequency such that it oscillates at frequency $\omega_i =
c_i + \wu_i$.  The purpose of the control system is to prevent the
elastic buffers from overflowing or underflowing, even though the
$\wu_i$ are not known exactly.  Full details of this mechanism are
presented in~\cite{bms}, where a mathematical model called the
\emph{abstract frame model}~(AFM) is developed.

Our focus in this paper is the use of proportional and proportional-integral control
for \bittide synchronization.  To that end we approximate the abstract
frame model with a simple linear model, removing the effects of
sampling and quantization.  We present simulations illustrating this
approximation, and analyze the mathematical properties
of the resulting linear system. We show that it is stable, and that
certain closed-loop performance metrics can be expressed in terms of
the resistance matrix of the graph. Performance is measured using the
$L_2$ norm of the buffer occupancies and frequency deviations, for
which we give exact formulae, in terms of the resistance distances in
the graph and the controller gains. These results directly relate the
connectivity of the graph to the performance properties of the \bittide
system.

\paragraph{Prior work.}
The synchronization mechanism of \bittide was first proposed
in~\cite{spalink_2006}. The abstract frame model for the system was
developed in~\cite{bms}, where a detailed description of the dynamic
behavior of the system was given. Another widely-used synchronous
network mechanism is SONET~\cite{sonet}.  The use of coupled-oscillators
to model synchronization originates with Winfree~\cite{winfree1967}.

In this paper, we approximate the \bittide mechanism with a linear
model. Our focus is on proportional-integral (PI) control, but the
corresponding model with purely proportional control is the widely
studied Laplacian dynamics, which has been extensively studied in the
literature, with applications including models of
flocking~\cite{boyds, jadbabaie}, Markov chain averaging
models~\cite{hastings,boyd2004}, congestion control
protocols~\cite{kelly}, power networks~\cite{strogatz,bullo}, vehicle
platooning~\cite{hedrick}, and consensus~\cite{olfati}.  Nonlinear
versions of Laplacian dynamics have been studied in~\cite{slotine},
and papers addressing PI control of Laplacian dynamics
include~\cite{burbano,freeman,andreasson}.

\section{Modeling}

We consider an oriented graph with $n$ nodes and $m$ edges.  Although
the graph is undirected, each edge has an \emph{orientation} used
purely to define the sign convention. For an oriented graph we define
the incidence matrix $B \in\R^{n \times m}$ by
\[
B_{il} = \begin{cases}
  1 & \text{if  node $i$ is the source of edge $l$} \\
  -1 & \text{if node $i$ is the target of edge $l$} \\
  0 &\text{otherwise}
\end{cases}
\]
and this defines a numbering of the edges $l=1,\dots,k$. The matrix
$L=BB^\tp$ is the \emph{Laplacian} matrix of the graph. We assume the
graph is connected. It is then a standard result that $L$ has rank
$n-1$.
There is exactly one zero eigenvalue,
with corresponding eigenvector $\one$.
See for example~\cite{godsil}. Choose $U_1$ to complete the
basis, so that  $U\in\R^{n\times n}$ is an
orthogonal matrix such that
\[
U = \bmat{U_1 & \one/\sqrt{n}}
\]
Then we can write $L$ in these coordinates so that
\[
U^\tp LU = \bmat{\hat{L} & 0 \\ 0 & 0}
\]
where $\hat{L} \in\R^{(n-1) \times (n-1)}$ is positive definite.

\paragraph{The abstract frame model.}
We briefly summarize the abstract frame model~\cite{bms}.
We have $n$ nodes.  At each node $i$ there is a clock,
whose value $\theta_i\in\R$ is called the \emph{clock phase}. We say
that $\theta_i$ measures local time at node $i$, in units called
\emph{local ticks.}  The rate of change for $\theta_i$ is called the
\emph{frequency} of node $i$, denoted by $\omega_i$. Every time $t$ at
which the phase $\theta_i(t)$ is an integer, node $i$ sends a data
frame to each of its neighbors.  The number of frames in the elastic
buffer at node $j$ associated with the link from node $i$ at time $t$ is
called the buffer \emph{occupancy}, denoted by $\beta_\itoj(t)$.  One
can show that
\[
\beta_\itoj(t)
 = \floor{\theta_i(t-l_\itoj)} - \floor{\theta_j(t)} + \ugn_\itoj
\]
where $\ugn_\itoj\in\Z$ is a constant, and $l_\itoj$ is the latency of the
link from $i$ to $j$.  Every $p$ local ticks, the controller at node
$i$ measures the buffer occupancies $\beta_\jtoi$ at that node. After a
delay of $d$ local ticks, the controller sets the frequency correction
$c_i$.  The delay parameter $d$ specifies the time required by the
controller to process the measurements and choose the frequency
correction and includes the time for the frequency change to take
effect on the physical oscillator.  The dynamic model is as follows.
For all $t \geq0$, $i \in \mathcal
V$, and~$k\in\Z_+$,
\begin{equation}
  \label{eqn:afm}
  \begin{aligned}
  \dot\theta_i(t) &= c_i^k + \wu_i \quad \text{for } t \in[s_i^k,s_i^{k+1}) \\
  \beta_\jtoi(t)   &= \floor{\theta_j(t-l_\jtoi)} - \floor{\theta_i(t)} + \ugn_\jtoi\\
  \theta_i(t_i^k) &= \theta_i^0 + kp  \\
  \theta_i(s_i^k) &= \theta_i^0 + kp + d \\
  y^k_i &= \{ (j, \beta_\jtoi(t_i^k)) \mid j \in \neighbors(i)\} \\
  \xi_i^{k+1} &= f_i^\text{disc}(\xi_i^k, y_i^k) \\
  c_i^k &= g_i^\text{disc}(\xi_i^k, y_i^k)
  \end{aligned}
\end{equation}
Because $\theta_i$ is an increasing function, the third and fourth equations
above uniquely determine the sampling times $t_i^k$ and the hold times $s_i^k$.
The controller is given by a discrete-time state-space system
$f_i^\text{disc}, g_i^\text{disc}$ at each node which maps the history
of these measurements to the correction~$c_i$.  The initial conditions
of the model are
\begin{equation}
  \label{eqn:ics}
  \theta_i(t) =
  \begin{cases}
    \theta_i^0 + \omega_i^{(-2)}t & \text{for } t \in [\epoch, 0] \\
    \theta_i^0 + \omega_i^{(-1)}t & \text{for } t \in [0, d/\omega_i^{(-1)}]
  \end{cases}
\end{equation}
These conditions are determined by initial frequencies
$\omega_i^{(-1)} > \omega^{\text{min}}$ and $\omega_i^{(-2)} >
\omega^{\text{min}}$, and initial clock phases $\theta^0_i \in\R^+
\backslash \Z$.  We are also given the initial buffer occupancies
$\beta^0_\jtoi \in\Z_+$, which together with~\eqref{eqn:ics} determine
the constants $\ugn_\jtoi$ such that
\[
\beta_\jtoi(t) - \beta^0_\jtoi = \floor{\theta_j(t-l_\jtoi)} - \floor{\theta_j(-l_\jtoi)}
- (\floor{\theta_i(t)} - \floor{\theta_i(0)})
\]
A controller is called \emph{admissible} if
\[
g_i^\text{disc}(\xi,y) + \wu_i >  \omega^{\text{min}}
\]
for all $i,k$ and all measurements $y$ and controller states~$\xi$. This ensures that
$  \omega_i(t)  > \omega^{\text{min}} \quad \text{for all }i, t$.
The time $\epoch<0$ is called the
\emph{epoch}, and it
must satisfy
\[
\epoch \leq -(l_\jtoi + d/\omega^\text{min}) \text{ for all } i,j \in \mathcal V
\]
We have shown in~\cite{bms} that the abstract frame model has a unique
solution under these conditions.

\subsection{An approximate model}

Our goal is to design a controller using a model for the system that
is as close to the AFM as possible. However, in order to
mathematically analyze and validate the controller, we need to use a
simplified model.  We perform two important simplifications.
Section~\ref{sec:pi} includes simulations of both the AFM and the
simplified model for comparison.

\paragraph{Continuous-time approximation.}
The first simplification is that we design a continuous-time
controller, instead of designing a discrete-time controller that uses
sampled-data.  Continuous-time control is not practically possible for
bittide, and therefore in implementations we need to discretize the
controller and subsequently validate that this does not adversely
affect performance.  With a sufficiently fast sampling rate, standard
approaches prove and demonstrate that this is an effective design
methodology.

We will use a continuous-time controller of the form
\begin{equation}
  \label{eqn:ctrl1}
  \begin{aligned}
    \frac{d \xi_i}{d\theta_i} &=  f_i(\xi_i, y_i) \\
    c_i(t) &=  g_i(\xi_i, y_i)
  \end{aligned}
\end{equation}
defined by functions $ f_i$ and $g_i$ at node $i$.  Notice here that
the independent variable is $\theta_i$, not $t$.  This captures the
dependence of the controller dynamics on the local oscillator
frequency, which arises because the clock that drives the
discrete-time controller is  provided by the oscillator at the
node. We write this in terms of $t$ as follows. Since $\dot \theta_i =
\omega_i$, we have
\begin{equation}
  \label{eqn:ctrl2}
  \begin{aligned}
    \dot \xi_i(t) &= \omega_i(t)  f_i(\xi_i(t), y_i(t)) \\
    c_i(t) &=  g_i(\xi_i(t), y_i(t))
  \end{aligned}
\end{equation}
The nodes in \bittide cannot exactly execute arbitrary dynamic control
since they do not have access to a perfect physical time reference.
Any integration or differentiation performed by the controller will be
scaled by the current clock frequency~$\omega_i$, which the controller
cannot measure.  Since the clock frequency is determined by the
controller, this introduces a nonlinear feedback into the system. The
magnitude of this effect depends on the range of frequency variation
experienced at the node. This is in practice determined by the
accuracy of the physical oscillators used. If the oscillator frequency
is accurate to within a relative error of $\alpha$, then only
correspondingly small relative corrections are required by the
controller, and therefore the term $\omega_i$ in
equation~\eqref{eqn:ctrl2} will be constant to within a relative error
$\alpha$ also. In a practical \bittide implementation we might see
$\alpha<10^{-5}$, which is substantially below the gain margin
typically used in control design.  Therefore this much uncertainty in
the controller parameters may be safely ignored. Hence we can
approximate the controller dynamics by
\begin{equation}
  \label{eqn:ctrl3}
  \begin{aligned}
    \dot \xi_i &= \wc  f_i(\xi_i, y_i) \\
    c_i(t) &=  g_i(\xi_i, y_i)
  \end{aligned}
\end{equation}
where $\wc$ is a constant approximation to the frequency (and hence not node specific.)

\paragraph{Quantization.}
The second simplification that we apply to the model is that we remove
the quantization of frames. That is, instead of enforcing the physical
property that the elastic buffer contains an integer number of frames,
we modify the model to allow the buffer occupancy to be non-integral. 
Replacing discrete-frames by a continuum results in a
so-called \emph{fluid model}, often used in analysis of stochastic
models of queuing systems~\cite{kurtz}. This corresponds to replacing
the expression for occupancy $\beta_\jtoi$ by
\[
\beta_\jtoi(t)
 = \bl(\floor{a\theta_j(t-l_\jtoi)} - \floor{a\theta_i(t)}\br)/a + \ugn_\jtoi
\]
and taking the limit as $a\to \infty$. Subject to mild
technical conditions, if the controller is linear, then solution
trajectories converge in $L_\infty$.  The limiting dynamics has an
approximate occupancy given by
\[
\beta_\jtoi(t) = \theta_j(t-l_\jtoi) - \theta_i(t) + \ugn_\jtoi
\]
We can expect this to be a good approximation if the frames are moving
sufficiently quickly through the system in comparison to the timescale
of the controller.

\paragraph{Zero delays.}
A further approximation that we make in this paper is that the
computation delays $d$ and the latencies $l_\jtoi$ are small enough to
be neglected, and set to zero. In practice this may or may not be the
case; some links have very long latencies. In other cases, such as
between machines in a datacenter, the latencies are very short.

The assumptions that quantization, loop delays, and sampling may be
ignored for the purposes of design, are well-studied in the literature
and frequently used in practice.  Mathematical techniques for handling
the error due to quantization
exist, for example~\cite{fu2005}. The problem of multi-rate
sampled-data systems has also been studied, for
example~\cite{mrsd,hassibi1999}. Systems with multiple delays are
analyzed in~\cite{mirkin}. In this case we have all of these phenomena
together with state dependent sampling rates and so further
analytical developments are required.
Thus, in this paper we do not validate the assumption that the error
induced by these approximations is small. We have as yet performed
neither the exhaustive numerical simulations nor the mathematical
analysis required to do so. However, we present some simple numerical
simulations below that indicate the approximation error is very small
in some cases of interest.

Combining these approximations and applying them to the abstract frame
model gives the following model:
\begin{equation}
  \label{eqn:simple}
  \begin{aligned}
    \dot\theta_i(t) &= c_i(t) + \wu_i \\
    \beta_\jtoi(t) - \beta^0_\jtoi &= \theta_j(t) - \theta_j^0
    - (\theta_i(t) - \theta_i^0)\\
  y_i(t) &= \{ (j, \beta_\jtoi(t)) \mid j \in \neighbors(i)\} \\
  \dot\xi_i(t) &= \wc f_i(\xi_i(t), y_i(t)) \\
  c_i(t) &= g_i(\xi_i(t), y_i(t))
  \end{aligned}
\end{equation}

\section{Controllers}

The objective of the control design is to ensure that the elastic
buffers neither overflow nor underflow. That is, we must ensure that
$0 \leq \beta_\jtoi(t) \leq \beta^\text{max}$
for all $t\geq 0$ and $i,j \in \mathcal V$.
The controller determines the frequency correction $c_i(t)$, and must
ensure that $\omega^\text{min} < c_i(t) + \wu_i < \omega^\text{max}$,
where $\omega^\text{min} > 0 $ and $\omega^\text{max}$ are the
physical limits of the oscillator.

To do this, we initialize the buffer occupancies in the middle of the
buffer, so that $\beta^0_\jtoi = \beta^\text{max}/2$ (
$\beta^\text{max}$ is even by construction.)  We will then construct a
feedback signal using the offset
\[
\bar\beta_\jtoi(t) = \beta_\jtoi(t) -\beta^0_\jtoi
\]
For convenience, we also define $\bar\theta_i(t) = \theta_i(t) - \theta_i^0$.
We will consider proportional and proportional-integral controllers. Since
each node may have a different number of neighbors, we must aggregate
the measurements of elastic buffer occupancy at each node, and we do
this simply by summing the occupancies, so that the controller at node $i$
uses the feedback signal
\[
r_i = \sum_{j \in \neighbors(i)} \bar\beta_\jtoi(t)
\]
Other choices for the measurement are possible. For example, one might
use the average buffer occupancy at a node, or the maximum occupancy
at a node. At this point in the analysis it seems as though any of
these choices might work well. However it turns out that using the sum
of occupancies results in the dynamics of the closed-loop system being
approximately equal to the well-known Laplacian dynamics on the graph,
which has many good properties, including guaranteed stability with
proportional control.

So far we have used notation $\beta\in\R^{n\times n}$ to denote
elastic buffer occupancy, with $\beta_\jtoi$ the occupancy of the
buffer at node $i$ associated with edge $(i,j)$, and $\beta_\itoj$ the
occupancy of the buffer at the other end of the edge. Under the
zero-latency assumption, we have $\bar\beta_\jtoi =  -\bar\beta_\itoj$.  We
therefore define the \emph{relative buffer occupancy}  $\delta \in\R^m$ by
\[
\delta_l = \bar\beta_\jtoi \text{ if $i$ is the source of edge $l$}
\]
and can replace $\bar\beta\in\R^{n\times n}$ by the equivalent
representation $\delta\in\R^m$.
We can now write the above dynamics as follows.
\begin{equation}
  \label{eqn:delta}
  \begin{aligned}
   \dot{\bar\theta}(t) &= c(t) + \wu \\
  \delta(t) &= -B^\tp \bar\theta(t)\\
  r(t) &= B \delta (t)\\
  \dot\xi_i(t) &= \wc f_i(\xi_i(t), r_i(t)) \\
  c_i(t) &= g_i(\xi_i(t), r_i(t))
  \end{aligned}
\end{equation}

\def\ts#1{\fboxsep=2pt\colorbox{white}{\small #1}}
\begin{figure*}[ht!]
  \centerline{\begin{overpic}[width=\textwidth]{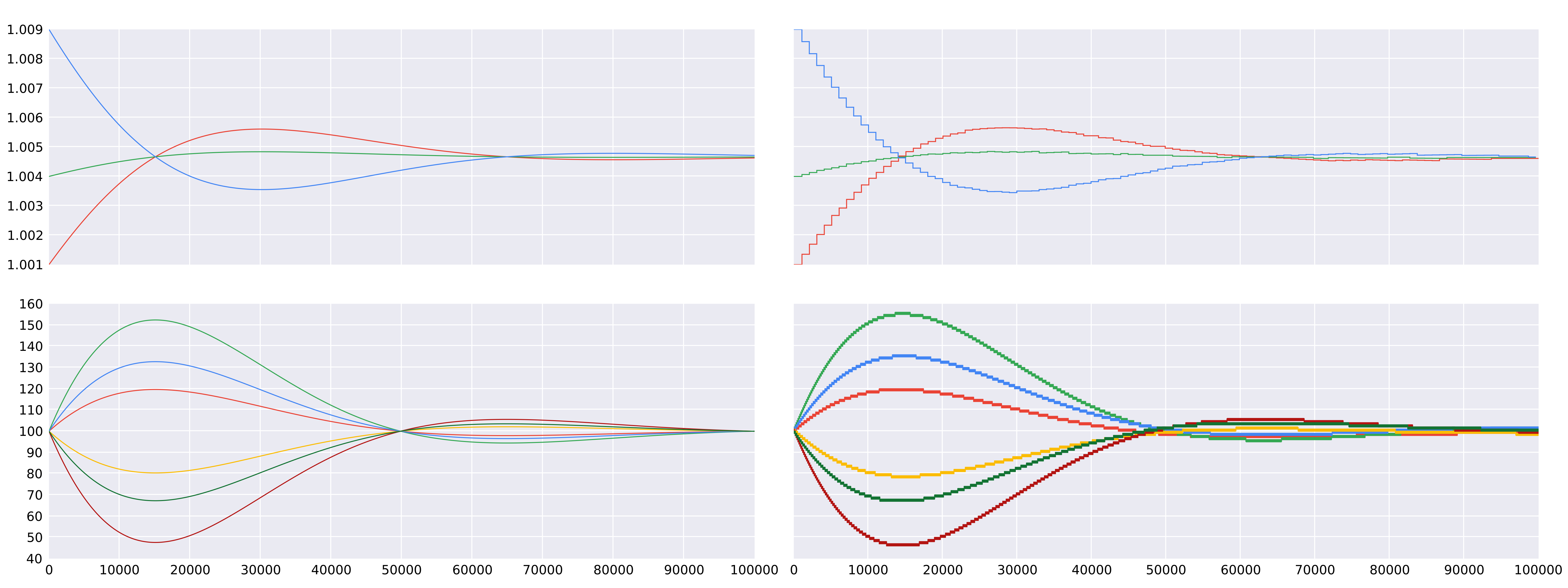}
      \put(30,32.5){\ts{ODE frequency $\omega$}}
      \put(80,32.5){\ts{AFM frequency $\omega$}}
      \put(30,15){\ts{ODE occupancy $\beta$}}
      \put(80,15){\ts{AFM occupancy $\beta$}}
      \put(45,3){\small $t$}
      \put(95,3){\small $t$}
    \end{overpic}}
    \caption{Comparison of trajectories from two different models}
  \label{fig:sims}
\end{figure*}
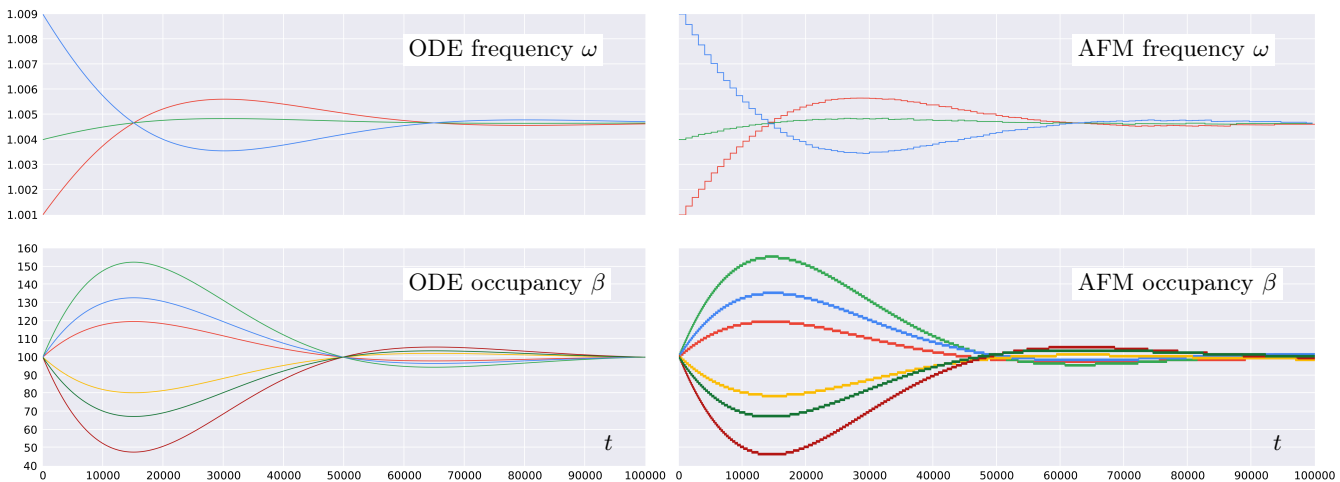

\section{Proportional-integral control}
\label{sec:pi}

We approximate a proportional-integral controller as follows.
The controller is
\begin{align*}
  \dot\xi(t) &= \wc r(t)  & \xi(0)&=0\\
  c(t) &= k_I \xi(t) + k_P r(t)
\end{align*}
Choosing state coordinates $x_1 = \bar \theta$ and $x_2 = \xi/\wc$,
we have, with  $x=(x_1,x_2)$, 
\begin{equation}
  \label{eqn:ode}
  \begin{aligned}
    \dot x &= A x + B_2 \wu \qquad x(0)=0\\
    \omega &= C_1 x + D_1 \wu \\
    \delta &= C_2 x
  \end{aligned}
\end{equation}
where for convenience $\kp = k_P$, $\wcki = \wc k_I$ and
\begin{align*}
  A &= \bmat{- \kp L & \wcki I \\
    - L & 0}
  &
  B_2 & = \bmat{ I \\ 0} \quad\\
  C_1 & = \bmat{-\kp L & \wcki I} &   D_1 &= I \\
  C_2 &= \bmat{-B^\tp & 0 }
\end{align*}

Figure~\ref{fig:sims} shows a comparison of the abstract frame
model~\eqref{eqn:afm} and the continuous-time ordinary differential
equation model~\eqref{eqn:ode}. This is for a graph with 3 nodes, with
bidirectional links connecting every pair of nodes. The system
parameters are $k_P = 3\times 10^{-5}$, $ k_I = 2\times 10^{-9}$,
$l_\jtoi = 500$, $p = 1000$, $d = 100$, $\theta^0_i = 0.1$,
$\omega^{(-2)} = \wu$, $\omega^{(-1)} = \wu$, and $\wc = 1$.  This
simulation shows that, for this choice of parameters, the
continuous-time model approximates the behavior of the AFM, at least
qualitatively. A more in-depth analysis of this approximation shows
that the effects of the quantization become insignificant for large
buffer occupancies.  As might be expected, the effects of the
latencies and computational delays are also negligible if the control
gains are not too large. We will not pursue this comparison further
here, but assume for the purposes of this paper that the parameters of
the system are such that the AFM behavior is well-approximated by the
differential equation model.

\section{System Behavior}
The proofs of the results in this section may be found in the Appendix.
Using the change of coordinates
\[
x = \bmat{
  U_1 & 0 & U_2 & 0 \\
  0 & U_1 & 0 & U_2} \hat x
\]
we have the dynamics
\begin{align*}
  \dot{\hat x}
  &= \bmat{
    -\kp \hat L & \wcki I & 0 & 0 \\
    -\hat L & 0 & 0 & 0 \\
    0 & 0 & 0 & \wcki I \\
    0 & 0 & 0 & 0 }
  \hat x
  +
  \bmat{U_1^\tp \\ 0 \\ U_2^\tp  \\ 0} \wu \\
  \omega
  &=
  \bmat{-\kp L U_1 & \wcki U_1 & 0 & \wcki U_2 } \hat x + \wu
  \\
  \delta &=
  \bmat{-B^\tp U_1 & 0 & 0 & 0} \hat x
\end{align*}
In these coordinates, we have immediately that $\hat x_4=0$,
since the initial conditions are $\hat x= 0$. This then implies
that
\[
\hat x_3(t) = n^\half \wavg t
\]
We now have the remaining dynamics
\begin{align*}
  \dot{\tilde x} &= \hat A \tilde x + \hat B_2 \wu \\
  \omega &= \hat C_1 \tilde x + \hat D_1 \wu \\
  \delta &= \hat C_2 \tilde x
\end{align*}
where $\tilde x= (\hat x_1, \hat x_2)$ and
\begin{equation}
  \label{eqn:hatsys}
  \begin{aligned}
    \hat A &= \bmat{-\kp \hat L & \wcki I \\
      - \hat L & 0}
    &
    \hat B_2 & = \bmat{ U_1^\tp \\ 0} \quad\\
    \hat C_1 & = \bmat{-\kp L U_1  & \wcki U_1} &   \hat D_1 &= I \\
    \hat C_2 &= \bmat{-B^\tp U_1  & 0 }
    &     \hat C &= \bmat{\hat C_1 \\ \hat C_2}
  \end{aligned}
\end{equation}
We would like to show that these dynamics are stable, which we state
formally here.
\begin{thm}
  \label{thm:stab}
  Suppose $a>0$ and $b>0$. Then the matrix $\hat A$  in~\eqref{eqn:hatsys} is Hurwitz.
\end{thm}
Since $\hat A$ is stable,  $\tilde x$ converges to a steady-state value
$\tilde x(t) \to \tilde x^\text{ss}$ as $t \to \infty$, and this is
\begin{equation}
  \label{eqn:ss}
  \tilde x^\text{ss} =  -\hat A^{-1}\hat B_2 \wu
  = \bmat{0 \\ \frac{1}{b} U_1^\tp} \wu
\end{equation}
We can now make the following observations regarding the behavior of
this system, which follow immediately from the above representation.

\begin{enumerate}[i)]
\item The phase $\bar \theta$ is the sum of a transient term
  which tends to zero and a linearly
  growing term whose growth rate is determined by the average
  frequency error. We have
  \begin{align*}
    \bar\theta(t)
    &= U_1 \hat x_1 + U_2 \hat x_3 \\
    &= U_1 \tilde x_1 +  \wavg t \one
  \end{align*}
  where $\tilde x_1 \to 0$ as $t \to \infty$.

\item The sum of the phases grows linearly with time, since
  $\one^\tp U_1 =0$.

\item The frequency $\omega = \dot{\bar\theta}$ has an invariant sum
  \[
  \sum_{i=1}^n \omega_i(t) = n \wavg
  \quad \text{ for all }t
  \]
  and all frequencies converge to the average frequency,
  that is $\omega_i(t) \to \wavg$ as $t \to \infty$.

\item The relative buffer occupancies tend to zero,
  that is $\delta(t) \to 0$ as $t \to \infty$.

\end{enumerate}

\subsection{Performance and Resistance Distance}

We now turn to performance measures of this controller. Specifically,
one of the primary controller objectives is to keep $\delta$ small.
This means that the buffer occupancies will remain close to the middle
of the buffer $\beta^0$, reducing the chance of the buffers
overflowing or underflowing. We consider here the \twonorm\ as a
measure of the magnitude of $\delta$. It will turn out that this
quantity is related to the connectivity graph of the system, and so this affords a
design strategy for the network topology. We can choose topologies
such that the norm of the buffer occupancy is small. However, we
note that this is simply a heuristic for the specific objective of
preventing buffer overflows and underflows.

For convenience let $\omega^\text{ss}$ be the steady-state frequency,
given by $\omega^\text{ss} = \omega^\text{avg} \one$.  We refer to the
quantity $\omega - \omega^\text{avg}\one$ as the \emph{frequency
deviation}, and the first result concerns its norm.

\begin{thm}
  \label{thm:freq}
  For the dynamics as above, we have
  \[
  \norm{\omega-\omega^\text{ss}}^2 = \frac{{\wu}^\tp L^\dag \wu}{2a}
  \]
\end{thm}

\noindent
The second result concerns the norm of the buffer occupancies.
\begin{thm}
  \label{thm:occ}
  For the dynamics as above, we have
  \[
  \norm{\delta}^2 = \frac{{\wu}^\tp L^\dag \wu}{2ab}
  \]
\end{thm}

\paragraph{The effect of the controller.} The above theorems separate
the effects on performance of the controller from the effects of the
graph. We see that increasing the proportional gain $\kp=k_P$ improves
both frequency and occupancy performance. However, increasing the
(scaled) integral gain $b=\wc k_I$ improves occupancy performance, but
does not change the norm of the frequency deviation.

\paragraph{The effect of the graph.}
The matrix $L^\dag$ is the pseudo-inverse of the Laplacian of the
graph, and has a well-known interpretation. Imagine a circuit
constructed according to the graph, with 1$\Omega$ resistors along
each edge.  Let $R_{ij}$ be the resistance of the resulting circuit
between nodes $i$ and $j$. This quantity is called the
\emph{resistance distance} of the graph.  Then
\[
R_{ij} = (e_i - e_j)^\tp L^\dag (e_i - e_j)
\]
where $e_i$ is the canonical basis vector~\cite{klein1993,gutman2004}.
This interpretation leads to several intuitive consequences. For
example, we have \emph{Rayleigh monotonicity}, the fact that adding an
edge cannot increase any $R_{ij}$. It is also immediately clear that
the resistance distance between any two nodes is less than or equal to
the path length between those nodes.

Theorems~\ref{thm:freq} and~\ref{thm:occ} show that the effect of the
graph on the $L_2$ performance of the \bittide system is entirely
through the matrix $L^\dag$ of resistance distances of the graph.  In
particular, we can see that \emph{adding edges can only improve
performance.} If we add an edge to the graph, then the resistance
distance between any pair of nodes cannot increase. Therefore, if the
norms of frequency deviation and relative buffer occupancy change,
they must decrease.

\subsection{Two Disequilibrated Frequencies}

An illustrative situation for a \bittide synchronization system is
when the system is almost in equilibrium, except for two
nodes. Consider a system in which there are two nodes, $i$ and $j$,
that have the frequencies $1+\alpha$ and $1-\alpha$, respectively. All
other nodes have uncorrected frequency~$1$.  That is
\[
\wu = \one + \alpha (e_i - e_j)
\]
Since $L^\dag\one = 0$, we have
\[
{\wu}^\tp L^\dag \wu = R_{ij}
\]
and using this, Theorems~\ref{thm:freq} and~\ref{thm:occ} give
the performance explicitly in terms of the resistance distance between
nodes $i$ and $j$ as
\[
\norm{\omega-\omega^\text{ss}}^2 = \frac{\alpha^2 R_{ij}}{2a}
\qquad
\norm{\delta}^2 = \frac{\alpha^2 R_{ij}}{2ab}
\]
Interpreting these results, we see that the norm performance for
both frequency and occupancy scales with the square root of the
resistance distance between the nodes.

\begin{figure}[ht!]
  \centerline{\relax
    \vcent{\hbox{\begin{overpic}[width=0.6\linewidth]{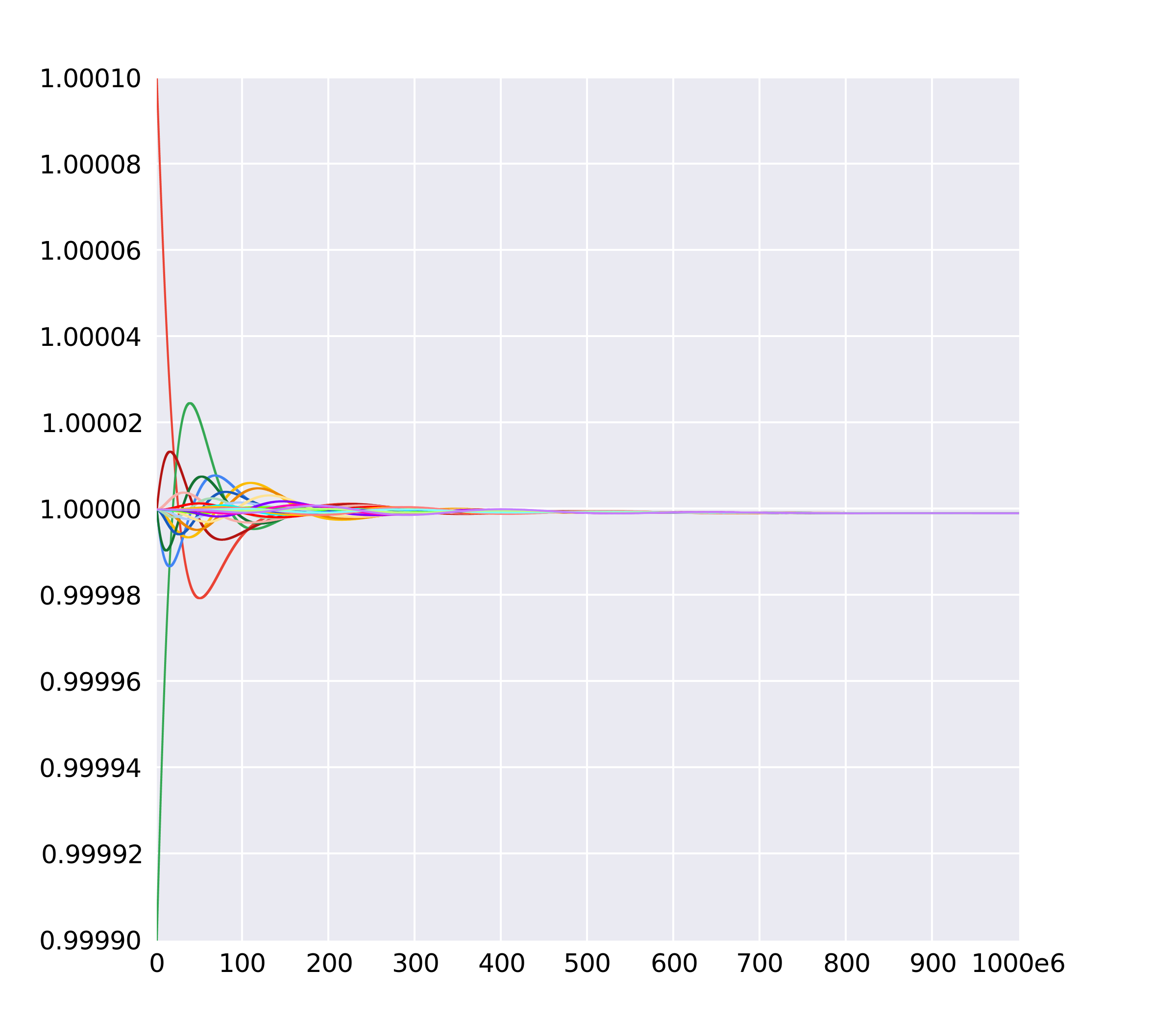}
      \put(0,40){\llap{\small$\omega$}}
      \put(50,-4){\clap{\small$t$}}
  \end{overpic}}}\relax
    \vcent{\hbox{\begin{overpic}[width=0.3\linewidth]{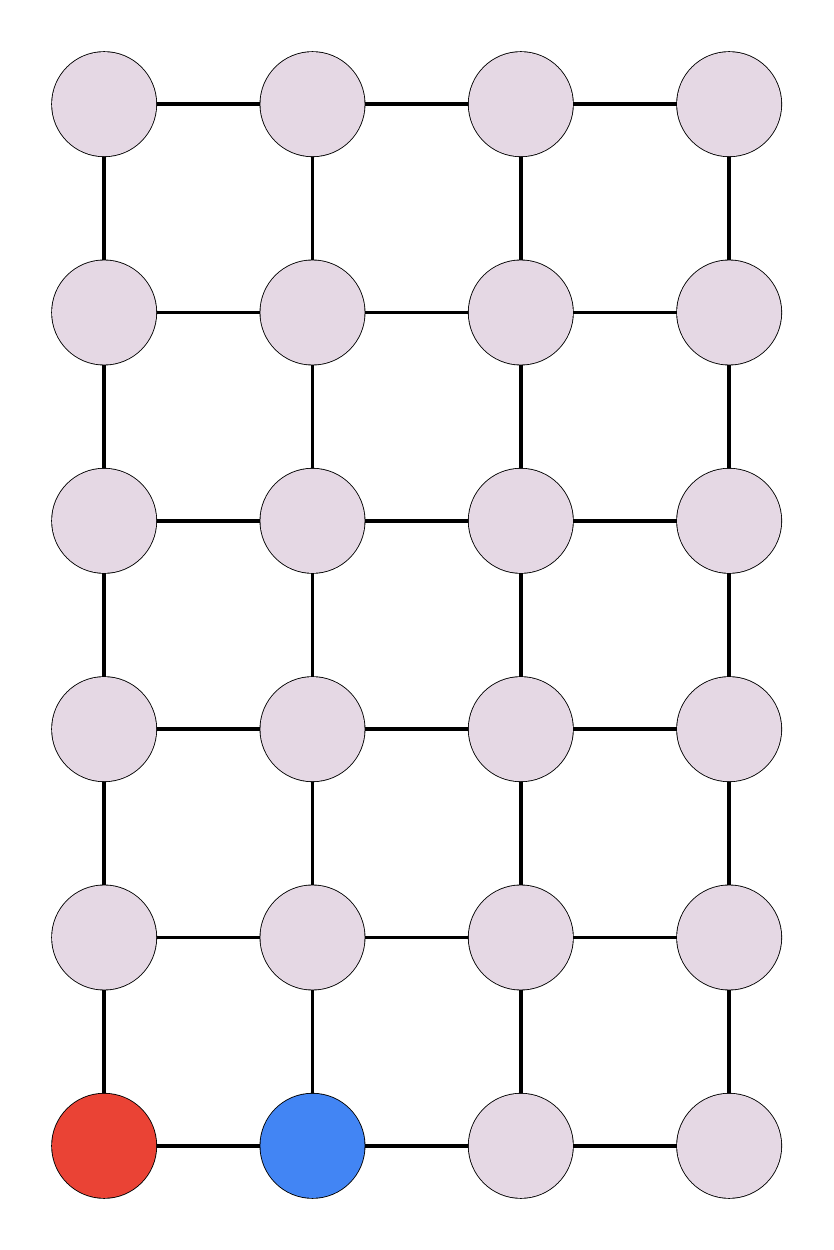}\end{overpic}}}}
  \precapspace
  \caption{Frequency for a system with two closely-spaced disequilibrated nodes}
  \label{fig:twoclose}
\end{figure}\begin{figure}[ht!]
  \centerline{\relax
    \vcent{\hbox{\begin{overpic}[width=0.6\linewidth]{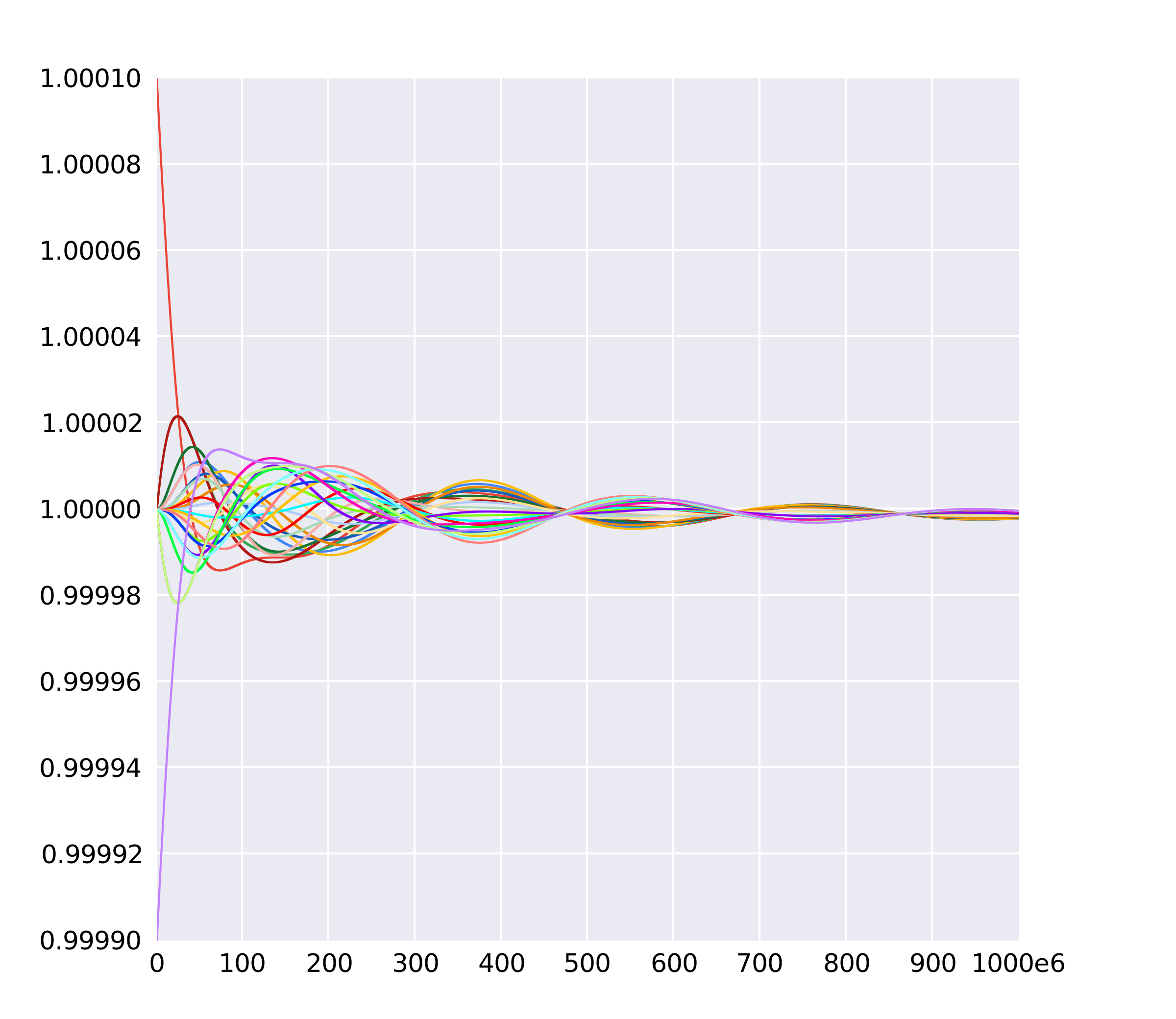}
      \put(0,40){\llap{\small$\omega$}}
      \put(50,-4){\clap{\small$t$}}
    \end{overpic}}}\relax
    \vcent{\hbox{\begin{overpic}[width=0.3\linewidth]{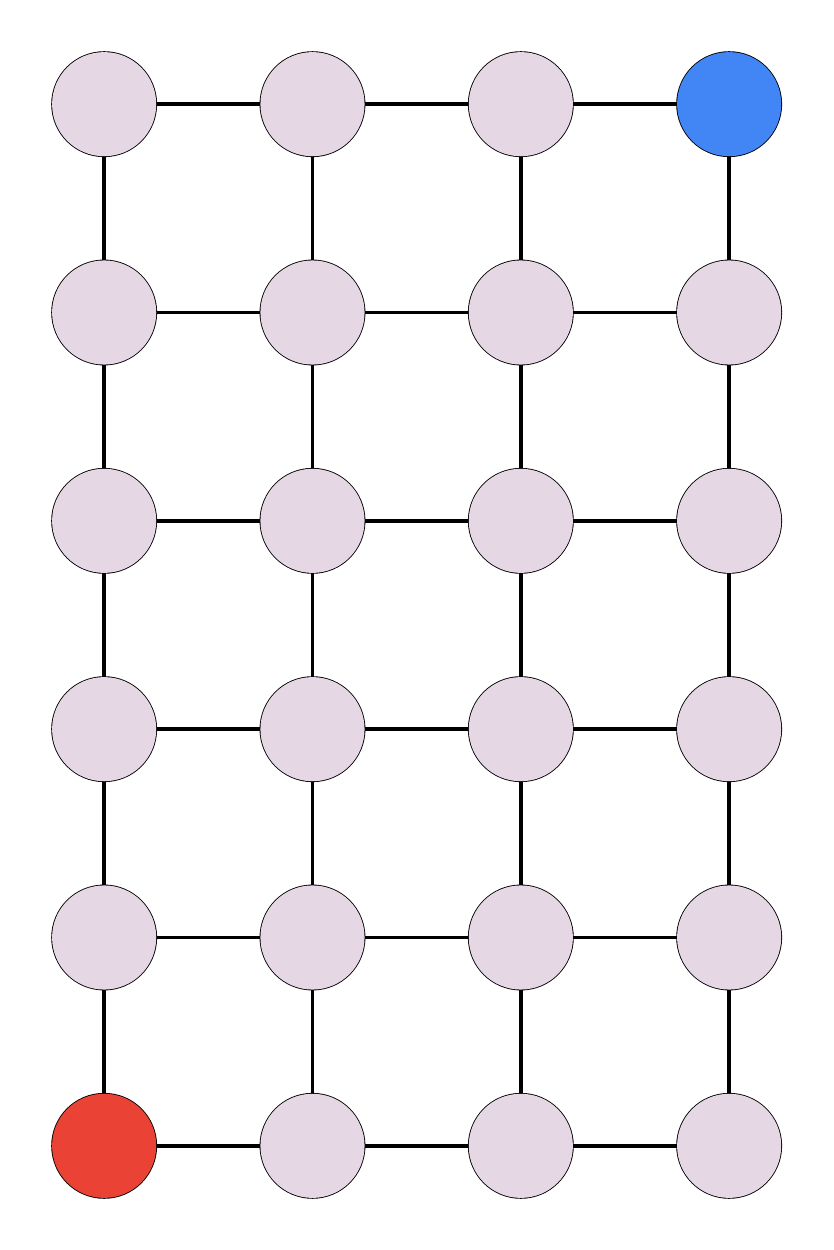}\end{overpic}}}}
  \precapspace
  \caption{Frequency for a system with two distant disequilibrated nodes}
  \label{fig:twofar}
\end{figure}

For example, consider a \bittide system on a $4\times 6$ mesh graph,
and let $\alpha=10^{-4}$. The parameters of this system are
$k_P = 2\times 10^{-8}$, $k_I = 10^{-15}$, $l_\jtoi = 5000$,
$p = 10^5$, $d = 1000$, $\theta^0_i = 0.1$,
$\omega^{(-2)} = \wu$,
$\omega^{(-1)} = \wu$, and $\wc = 1$.
We consider two examples, one in
Figure~\ref{fig:twoclose} where nodes~$i$ and $j$ have small
resistance distance $R\approx 0.700$, and another in
Figure~\ref{fig:twofar} where $R \approx 2.262$.
In both figures the graph is shown with node $i$ highlighted
in red and node $j$ highlighted in blue. As a consequence, we have for
Figure~\ref{fig:twoclose}
\[
\norm{\omega-\omega^\text{ss}}^2 \approx 0.175
\]
and for Figure~\ref{fig:twofar}
\[
\norm{\omega-\omega^\text{ss}}^2 \approx 0.565
\]
The greater resistance between the perturbed nodes in
Figure~\ref{fig:twofar} leads to worse performance,
as shown by the slower convergence of frequency divergence
in the figure.

\subsection{Worse-Case Frequencies}

Another application of Theorems~\ref{thm:freq} and~\ref{thm:occ} is
that they allow computation of the worst-case uncontrolled frequency
distribution $\wu$.  The norm response of both occupancy and frequency
deviation is proportional to
\[
f(\wu) = {\wu}^\tp L^\dag \wu
\]
We consider all uncontrolled frequencies such that
\[
\norm{\wu}_2\leq \gamma
\]
and seek to maximize $f(\wu)$ over this bounded set. Bounding the set
is important for the problem to be mathematically meaningful, since
otherwise we can make $f(\wu)$ large simply by scaling $\wu$. However,
the results here are determined also by the way in which we have
chosen to bound $\wu$. Here we choose the Euclidean norm primarily
because for this choice we can compute exactly the maximum of $f$ and
the corresponding worst-case $\wu$. Such a choice could be motivated
by assuming a Gaussian probabilistic model for $\wu$. Alternatively, a
deterministic formulation might be better suited to an analysis using
the $\infty$-norm. We do not delve further into these alternatives
here, but instead view the choice of set as a rough proxy for a more
accurate model of the set of possible uncontrolled frequencies.

The $x$ that maximize a homogeneous quadratic function $x^\tp Q x$
over $x\in\R^n$ with $\norm{x}\leq 1$ is given by $x= v$, where $v$ is
the unit eigenvector of $Q$ with largest eigenvalue.  Here we consider $Q=
L^\dag$.
 The second smallest eigenvalue of $L$ is called
the \emph{algebraic connectivity} of the graph, and the corresponding
unit eigenvector is called the \emph{Fiedler vector}~\cite{fiedler}.
Since exactly one eigenvalue of $L$ is zero, and the others are
strictly positive, the eigenvector $v$ that maximizes $x^\tp Q x$
is the Fiedler vector.

\begin{figure}[ht!]
  \centerline{\vcent{\hbox{\begin{overpic}[scale=0.16]{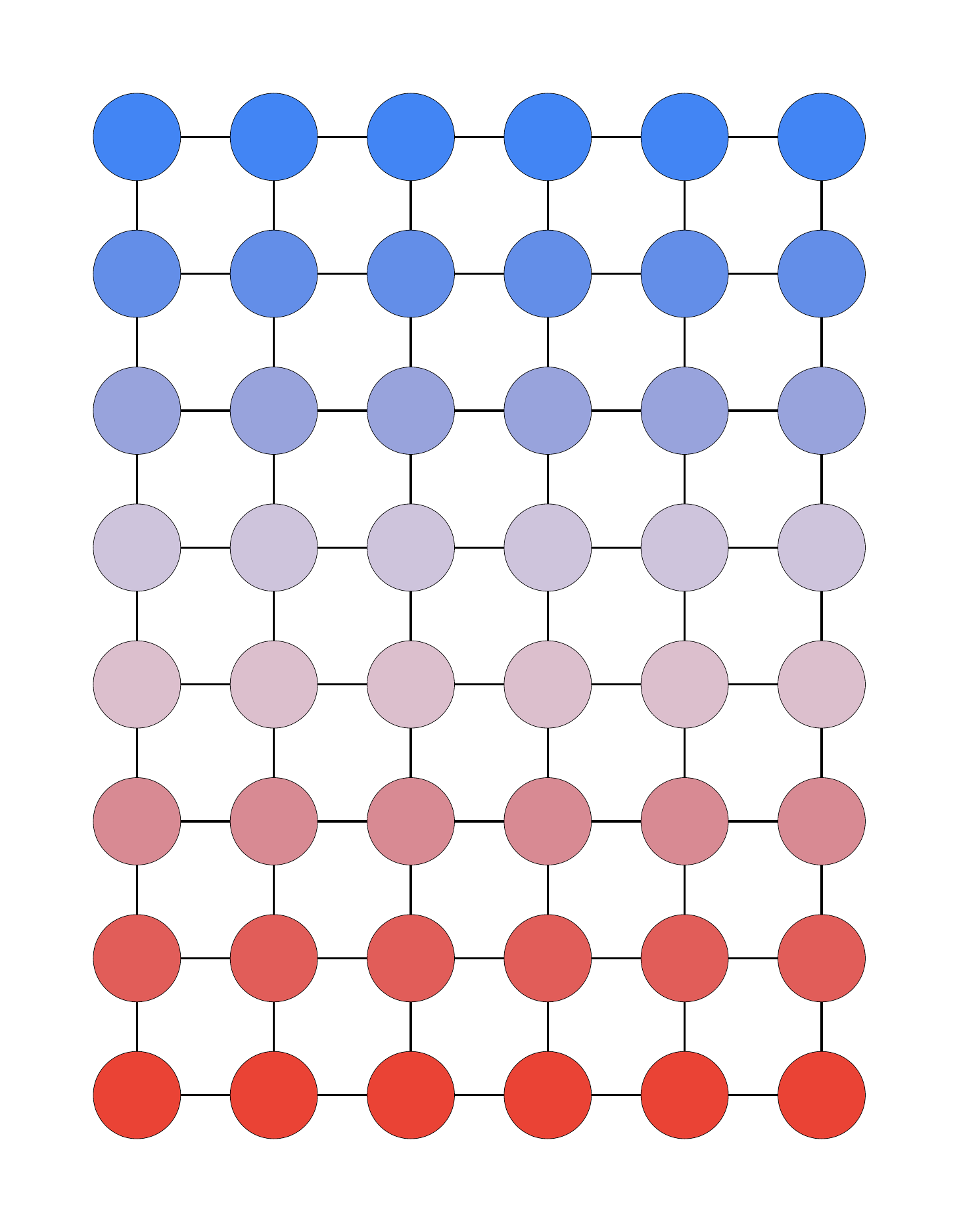}\end{overpic}}}\hss
    \vcent{\hbox{\begin{overpic}[scale=0.16]{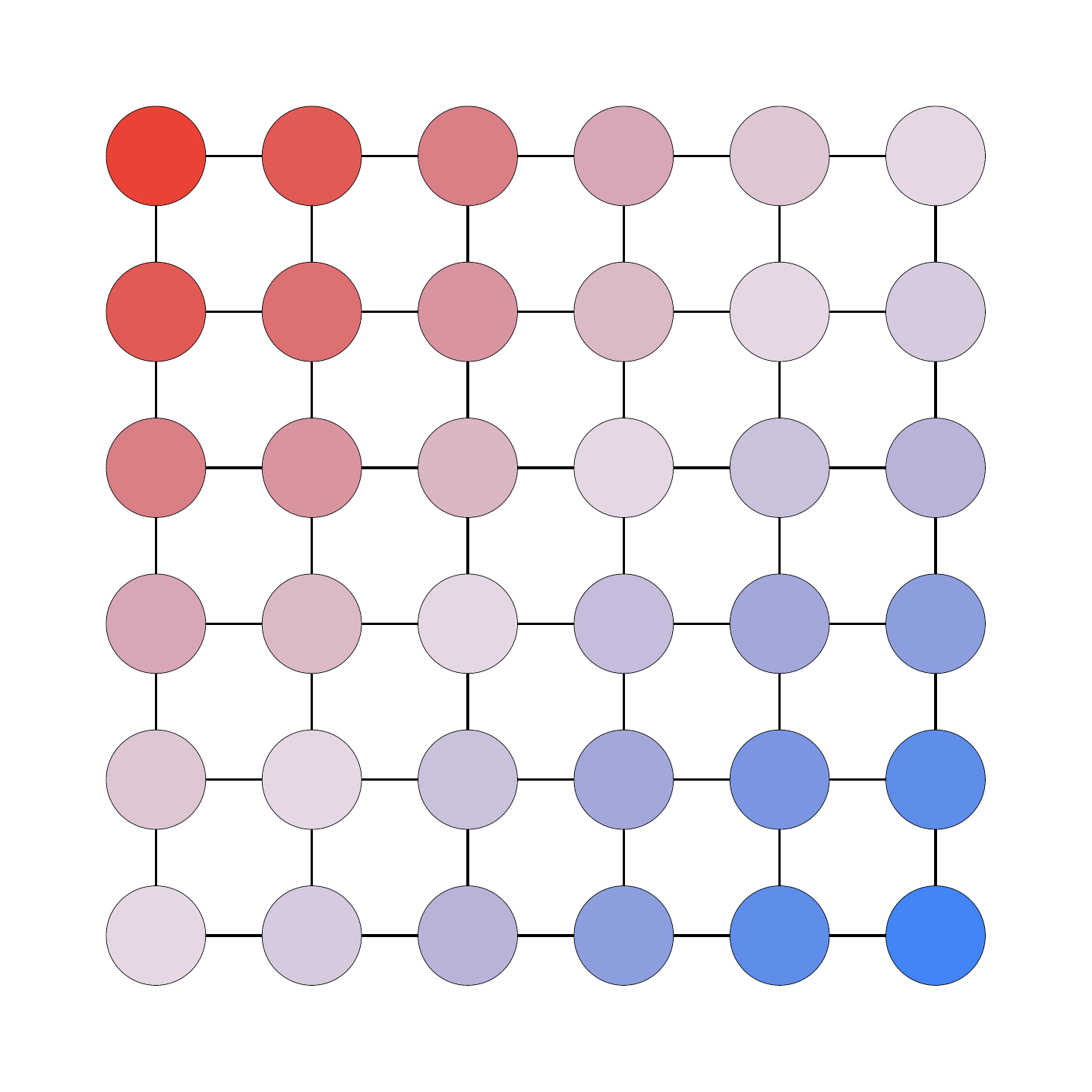}\end{overpic}}}\hss
    \vcent{\hbox{\begin{overpic}[scale=0.16]{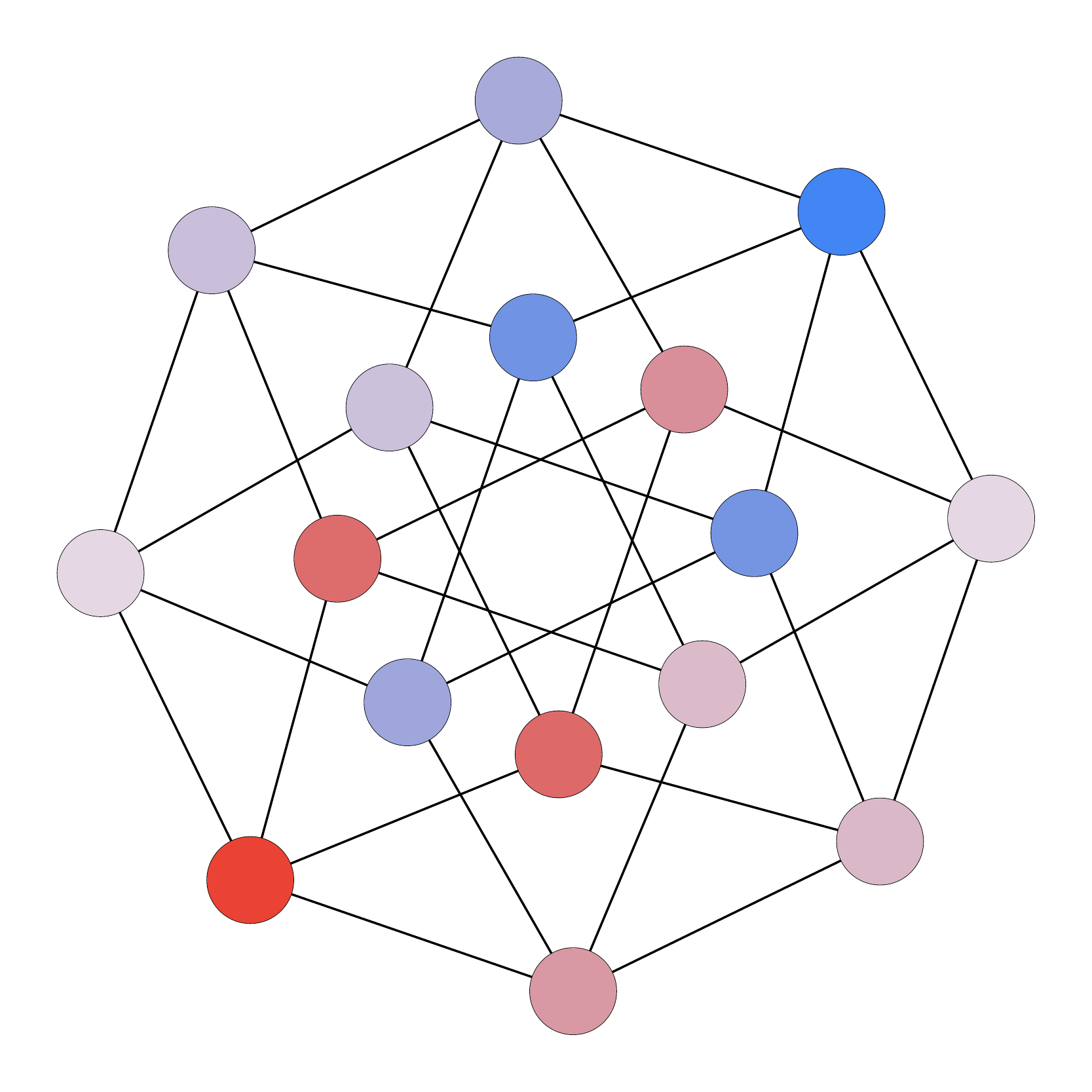}\end{overpic}}}}
  \caption{Worst-case uncontrolled frequencies}
  \label{fig:wcuf}
\end{figure}

Figure~\ref{fig:wcuf} shows the corresponding worst-case frequency
distributions for three example graphs. Here red shows positive values
of $\wc_i$ and blue shows negative values; the exact scale is omitted
since it is an arbitrary consequence of the magnitude of $\wc$.
Notice that for the rectangular grid graph shown, the worst-case
distribution varies from top-to-bottom, whereas for the square grid it
varies diagonally, even though the greatest resistance between any two
nodes on the rectangular graph is achieved by diagonally opposite corners.

\section{Conclusions}

In this paper we have analyzed the performance of \bittide
synchronization using the $2$-norm of the frequency deviation and
relative buffer occupancies. When using PI control, we have shown
these quantities are determined in a simple way by the
control gains and by the graph resistances. We used these results
to analyze and illustrate some simple examples, showing the utility
of resistance as a performance indicator in these systems.

\section*{Acknowledgments}

We thank Jean-Jacques Slotine for initial guidance on controller
behavior. We thank Sahil Hasan and Tong Shen for all their work on the
project.

\section*{Appendix: Proofs}

Consider the dynamics $\dot  x =  A x$, with output $y=Cx$ and Lyapunov
equation
\[
A^\tp X + XA + C^\tp C = 0
\]
We will need the following two standard results from linear systems theory.
\begin{lem}[Theorem 4.1 in~\cite{dullerud}]
  \label{lem:lyapperf}
  If $A$ is Hurwitz, then the Lyapunov equation has a
  unique solution $X\in\R^{n\times n}$ and $\norm{y}_2^2 = x(0)^\tp X x(0)$.
\end{lem}
\begin{lem}[Proposition 4.2 in~\cite{dullerud}]
  \label{lem:lyapstab}
  If $C^\tp C >0$ and $X>0$ satisfies the Lyapunov equation, then $A$
  is Hurwitz.
\end{lem}

First, to reduce the system~\eqref{eqn:hatsys} to this form, we need
to remove the non-zero limiting value of the state, which is induced
by the constant forcing term $\wu$.
Define $\bar x = \tilde{x} -\tilde{x}^\text{ss}$
and $\bar \omega = \omega - \omega^\text{ss}$ then we have dynamics
\begin{align*}
  \dot{\bar{x}} &= \hat A \bar x   & \bar{x}(0) = \hat{A}^{-1}\hat B_2 \wu \\
  \bar\omega &= \hat{C}_1 \bar x \\
  \delta &= \hat{C}_2 \bar x
\end{align*}
Now we can solve the Lyapunov equations.
Define
\[
X_1 = \bmat{ \frac{a}{2} \hat L + \frac{b}{2 a}I  &
  -\frac{b}{2}I \\[1mm]
  -\frac{b}{2}I &
  \frac{b^2}{2a} \hat{L}^{-1}}
\quad
X_2 = \bmat{ \frac{1}{2 a} I & 0 \\[1mm]
  0 & \frac{b}{2 a} \hat L^{-1}}
\]

\noindent\textbf{Proof of Theorem~\ref{thm:stab}.}
First we show that $\Null(\hat C) = \{0\}$. This follows because
\[
\hat C = \bmat{-a L U_1 & bU_1 \\ -B^\tp U_1 & 0}
\]
The $1,2$ block satisfies $\Null(U_1)=\{0\}$ since $U_1$ has
orthonormal columns. The $2,1$ block satisfies
\[
\Null (B^\tp U_1) = \Null(U_1^\tp BB^\tp U_1) = \Null(\hat L) = \{0\}
\]
Therefore $\hat C^\tp \hat C > 0$. Now one can verify that with $X=X_1
+ X_2$, we have $\hat A^\tp X + X \hat A + \hat C^\tp \hat C = 0$.
Finally, $X_1>0$ (via the Schur complement condition) and $X_2>0$,
and hence by Lemma~\ref{lem:lyapstab} the matrix $\hat A$ is Hurwitz. \hfill\qed

\noindent\textbf{Proof of Theorem~\ref{thm:freq}.}
One can verify that $\hat A^\tp X_1 + X_1 \hat A + \hat C_1^\tp  \hat C_1 =0$.
Using Theorem~\ref{thm:stab} we know that $\hat A$ is Hurwitz, and
hence Lemma~\ref{lem:lyapperf} implies that
\begin{align*}
  \norm{\bar w}^2
  &= \bl(\hat{A}^{-1}\hat B_2 \wu\br)^\tp X_1  \hat{A}^{-1}\hat B_2 \wu \\
  &= \frac{1}{2a} {\wu}^\tp L^\dag \wu
\end{align*}
as desired. \hfill\qed

\noindent\textbf{Proof of Theorem~\ref{thm:occ}.}
One can verify that $\hat A^\tp X_2 + X_2 \hat A + \hat C_2^\tp  \hat C_2 =0$.
Using Theorem~\ref{thm:stab} we know that $\hat A$ is Hurwitz, and
hence Lemma~\ref{lem:lyapperf} implies that
\begin{align*}
  \norm{\delta}^2
  &= \bl(\hat{A}^{-1}\hat B_2 \wu\br)^\tp X_2  \hat{A}^{-1}\hat B_2 \wu \\
  &= \frac{1}{2ab} {\wu}^\tp L^\dag \wu
\end{align*}
as desired. \hfill\qed

\prerefspace

\end{document}